\documentclass[a4paper]{IEEEtran}

\usepackage[hidelinks,bookmarks=false]{hyperref}
\usepackage[official]{eurosym}
\usepackage{flushend,amsmath}
\usepackage{array}

\usepackage[utf8]{inputenc}

\ifCLASSINFOpdf
  \usepackage[pdftex]{graphicx}
  \graphicspath{{graphics/}}
  \DeclareGraphicsExtensions{.pdf,.jpeg,.png}
\else
\fi

\usepackage{tikz}
\usepackage{url}

\usepackage[europeanresistors,americaninductors]{circuitikz}
\usepackage{adjustbox}

\usepackage{booktabs}

\begin{document}
%

\title{Modeling Curtailment in Germany:\\How Spatial Resolution Impacts Line Congestion}

\author{\IEEEauthorblockN{Martha Frysztacki, Tom Brown}\\ \vspace*{.1cm}
  \IEEEauthorblockA{Institute for Automation and Applied Informatics, Karlsruhe Institute of Technology, 76344~Eggenstein-Leopoldshafen, Germany \\ \vspace*{.1cm}
e-mail: \href{mailto: martha.frysztacki@kit.edu}{martha.frysztacki@kit.edu}}
}


%


\maketitle
\IEEEpubid{978-1-7281-6919-4/20/$\$$31.00 ©2020 IEEE}
\IEEEpubidadjcol

\begin{abstract}
	This paper investigates the effects of network constraints in energy system models at transmission level on renewable energy generation and curtailment as the network is being spatially aggregated. We seek to reproduce historically measured curtailment in Germany for the years 2013-2018 using an open model of the transmission system, PyPSA-Eur. Our simulations include spatial and temporal considerations, including congestion per line as well as curtailment per control zone and quarter.
		
	Results indicate that curtailment at high network resolution is significantly overestimated due to inaccurate allocation of electricity demand and renewable capacities to overloaded sites. However, high congestion rates of the transmission network decrease as the network is clustered to a smaller number of nodes, thus reducing curtailment. A measure to capture errors in the assignment of electricity demand and power plants is defined and hints towards a preferable spatial resolution. Thus, we are able to balance the effects of accurate node assignment and network congestion revealing that a reduced model can capture curtailment from recent historical data. This shows that it is possible to reduce the network to improve computation times and capture the most important effects of network constraints on variable renewable energy feed-in at the same time.
\end{abstract}
\textbf{keywords}: curtailment, spatial clustering, energy system modeling at transmission level, renewable energy


%
\IEEEpeerreviewmaketitle

\section{Introduction}
To substantially reduce the risks and impacts of climate change, the European Council has committed to becoming climate neutral by 2050 \cite{Roadmap2050}. This objective is at the heart of the European Green Deal \cite{GreenDeal} and in line with the Paris Agreement, that sets out the goal to pursue efforts to limit global warming to 1.5$^\circ$C \cite{Paris}.

The incentives for renewable energy in Germany are regulated by the Renewable Energy Sources Act. If the feed-in of electricity from an installation to generate electricity from renewable energy sources is reduced due to a grid system bottleneck, grid system operators must compensate operators affected by their measure for $95\%$ of the lost revenues \cite{EEG2017}. The compensation payments for the curtailed energy in 2013 were $43.7$ million euros and have increased ever since, reaching a maximum of $718.7$ million euros in 2018 \cite{EEG18, Monitoring19}. The reason is that in the last few years significant amounts of wind have been curtailed because of congestion in the German transmission system. 

At the same time, the German government has set a target that the share of renewables must be increased from $40\%$ in 2019 to $65\%$ by 2030. By 2050 the capacities of today must at least quintuple even in optimistic scenarios to meet CO$_2$ reduction targets \cite{ISE2050}. Thus, congestion is likely to continue to be present as shares of wind and solar rise, particularly given the delays in building new transmission projects. Various solutions have been proposed to mitigate congestion: flexibility from sector-coupling \cite{Brown2018}, the production of green hydrogen \cite{Wasserstoffstrategie}, innovative new technologies such as dynamic line rating \cite{Wallnerstroem2015} or fast-acting storage \cite{Netz19} could be introduced to the market to increase available green energy from existing assets.

Allowing new flexibility strategies to be tested on openly-available, validated models should increase transparency and innovation in managing congestion \cite{Pfenninger2018}. Therefore, accurate modelling of the interactions between renewables and the grid is critical to assess future scenarios for the energy system.

This work contributes to the literature by validating the historical
curtailment in the German network in an open model of the European power
system, PyPSA-Eur, which is free to download, modify and republish. Model adaptations and measures are introduced to explore how best to model the curtailment while not overloading
scarce computational resources. In particular we vary the spatial
resolution of the model to understand at what level the most important
bottlenecks are still captured, balancing this against the overall size
of the model. Results are validated against published feed-in management numbers by the German Federal Network Agency \cite{BNetzA18, BNetzA19} in time and space.

This paper is arranged as follows: In Section \ref{sec:modelandmethods} we present the full optimisation problem with respect to all its constraints to model curtailment in Germany. We provide the data inputs and present the applied clustering methodology based on k-means. In Section \ref{sec:results}, we present our results of clustering on curtailment. Based on these findings, model results are discussed on an annual scale for the years 2013-2018, a spatial scale for the four distinct TSO regions in Germany and on a temporal scale discussing results per quarter.

\IEEEpubidadjcol

\section{Model and Methods} \label{sec:modelandmethods}
\begin{table}\centering
	\caption{Notation} \label{tab:notation}
	\begin{tabular}{@{} lp{5cm} @{}}
		\hline
		abbrev. & description\\
		\hline
		& general abbreviations \\
		\hline
		$s$ & technology type (generators/sorage) \\
		$t$ & time discretisation\\
		$a$ & year \\
		$n$ & substation/node \\
		$\ell$ & transmission line$^1$\\
		$N_{c}$ & set of high resolution nodes in cluster $c$ \\
		\hline
		& line attributes \\
		\hline
		$x_\ell $ & reactance\\
		$l_\ell$ & length \\
		$F_\ell^a$ & capacity in year $a$\\
		$f_{\ell,t}$ & energy flow \\
		\hline
		& nodal attributes \\
		\hline
		$\mathrm{gdp}_n$ & gross domestic product in node $n$ \\
		$\mathrm{pop}_n$ & population in node $n$ \\
		$x_n, y_n$ & coordinates of node $n$ \\
		$G_{n,s}^a$ & capacity in year $a$\\
		$o_{n,s}$ & variable costs\\
		$\eta$ & storage losses or efficiencies for technology\\
		$d_t$ & demant in time $t$ (whole Germany) \\
		$d_{n,t}$ & demand per node $n$ and time $t$ \\
		$\bar{g}_{n,s,t}$ & capacity factor for RE, $\in [0,1]$\\
		$g_{n,s,t}$ & dispatch\\
		$A_{n,s,t}$ & availability of renewables in TWh \\
		\hline
		& graph related attributes \\
		\hline
		$K_{n,\ell}^a$ & incidence matrix in year $a$ \\
		$C_{\ell,c}^a$ & cycle matrix in year $a$; $c$ represents a cylce\\
		\hline
	\end{tabular}\\
	\flushleft{$^1$ Usually the context is clear, so the indices accounting for the nodes $n,m$ connected by line $\ell_{n,m}$ are omitted for simplicity reasons.}
\end{table}

\subsection{Optimisation problem}
While PyPSA-Eur \cite{PyPSA, PyPSA-Eur} is capable of co-optimising investment in generation and transmission, this research is a case-study to reproduce historical data. Therefore, the objective function minimises solely the operational costs for a fixed generation and storage fleet and fixed transmission capacities. The historic capacities of the generation and transmission fleet are given exogenously for each year, labeled by an index $a$.
\begin{align} \label{eq:systemcosts}
\min_{\substack{g_{i,s,t},\ f_{\ell,t}}} \Bigl[ \sum_{n,s,t} o_{i,s}g_{i,s,t} \Bigr] \,.
\end{align}

The objective to minimise operational expenditures is constrained to satisfy the electricity demand $d_{n,t}$ at each node $n$ and in each time $t$, either by local generators and storage options or by the energy flow $f_{\ell,t}$ from neighboring nodes:
\begin{align} \label{eq:KCL}
\sum_{n,s,t} g_{n,s,t}-d_{n,t} = \sum_\ell K_{n,\ell}^a f_{\ell,t}\quad \forall n,t \,,
\end{align}
where $K_{\ell,t}^a$ is the incidence matrix of the network, encoding its topology for each year $a$, to reflect connections between nodes. This equation represents Kirchhoff's Current law, stating that the sum of currents flowing into a node must equal the sum of out-flowing currents minus local consumption. The transmission grid is additionally constrained by Kirchhoff's voltage law, stating that the directed sum of potential differences around any closed cycle adds to zero. This can be translated to direct constraints on the flows \cite{Hoersch2018}:
\begin{align} \label{eq:KVL}
\sum_\ell C_{\ell,c}^a x_\ell f_{\ell,t} = 0 \quad \forall c,t \,.
\end{align}
Each cycle $c$ is represented in the matrix $C_{\ell,c}^a$ as a directed combination of lines $\ell$. $x_\ell$ denotes the inductive reactance.

Further, each flow in the transmission grid is constrained by the line capacity multiplied by a factor of $0.7$, a convention used in the past decades to account for $N-1$ security, see \cite{(n-1)-1, (n-1)-2}:
\begin{align} \label{eq:linecapacity-constraint}
| f_{\ell,t} | \leq 0.7\cdot F_{\ell}^a \quad \leftrightarrow \quad
\left\{ \ \begin{matrix}
\overline{\mu}_{\ell,t}^a \geq 0 \\
\underline{\mu}_{\ell,t}^a \geq 0
\end{matrix}\right. \qquad \forall \ell,t\,.
\end{align}
The shadow prices $\overline{\mu}_{\ell,t}^a$ and $\underline{\mu}_{\ell,t}^a$ [\euro/MW] are positive, if the flow $f_{\ell,t}$ equals $70\%$ of its capacity, and if this constraint is binding, i.e. a better optimum of the overall annual costs according to the objective in (\ref{eq:systemcosts}) could be reached by increasing $f_{\ell,t}$ beyond this constraint.

The dispatch of conventional generators $g_{n,s,t}$ is constrained by their given capacities $G_{n,s}^a$
\begin{align} \label{eq:generator-constraint}
0 \leq g_{n,s,t} \leq G_{n,s}^a \quad \forall n,s,t\,.
\end{align}
In case of renewables, an additional availability factor $\bar{g}_{n,s,t} \in [0,1]$ is added to the same constraint to reflect the spatio-temporal variability of weather conditions:
\begin{align} \label{eq:RE-constraint}
0 \leq g_{n,s,t} \leq \bar{g}_{n,s,t} \cdot G_{n,s}^a \quad \forall n,s,t\,.
\end{align}


The energy levels $e_{n,s,t}$ of all storage units (we only include hydro storage) have to be consistent, i.e. the current storage level equals the previous storage level plus what is charged and discharged, accounting for standing, charging and discharging efficiencies. The energy level is assumed to be cyclic, such that by the end of the simulated year on December 31$^{st}$ the storage is filled by the same amount as it was assumed on January 1$^{st}$.

\subsection{Data Inputs}

We model Germany with a maximum of 306 nodes, including all transmission lines from the ENTSO-E Interactive Transmission Map \cite{ENTSO-E} extracted by the GridKit extraction tool \cite{GridKit}. The network is adjusted according to annual reports from local transmission system operators \cite{Netz13}-\cite{Netz19}. Lines that were not build by the time of e.g. 2014, are removed from the optimisation for this year. Lines that have been strengthened were reduced in capacity for the optimisation. Electricity demand data \cite{OPSD2019} and generation time series for hydroelectricity (run of river) are included and fixed. Capacities for conventional and renewable generators are taken from the
new database provided by the German Federal Network Agency, the
Marktstammdatenregister \cite{MaStR2019}, see Figure \ref{fig:germanymap} for their spatial distribution. Fuel costs, variable operation and maintenance costs per technology are based on historical market prices \cite{DIW}. Renewables have no marginal costs, but were given very small ones to set the curtailment order for wind and solar: 0 ct/MWh$_\mathrm{el}$ for run of river and geothermal, 1 ct/MWh$_\mathrm{el}$ for solar, 2 ct/MWh$_\mathrm{el}$ for wind onshore and 3 ct/MWh$_\mathrm{el}$ for wind offshore.

Electricity-Demand in \cite{OPSD2019} is given per country in hourly resolution $d_t$; therefore, we need to dis-aggregate it in space over Germany. We do this by applying a heuristic based on the gross domestic product $\mathrm{gdp}_n$ and population $\mathrm{pop}_n$ per node $n$ based on NUTS3 data \cite{NUTS3}:
\begin{align} \label{eq:loadassignment}
	d_{n,t} = {d_t} \cdot \left( 0.6 \cdot \|\mathrm{gdp}_n\|_\mathrm{max} + 0.4 \cdot  \|\mathrm{pop}_n\|_\mathrm{max} \right) \,.
\end{align}

A similar heuristic was evaluated in \cite{Zhou2005} and matches the electricity demand fairly well with a small assignment error.

The dataset \cite{MaStR2019} contains the geographic coordinates or equivalent information, i.e. each generator $g$ lies within a so-called \textit{voronoi cell}. These cells are defined by a center point and cover the space that is closest in the sense of the euclidean metric. Therefore, we assign each generator $g$ to its closest cell, that is represented by the node $n$ of the network via
\begin{align} \label{eq:generatorassignment}
\mathrm{argmin}_{n\in\mathcal{N}} \sqrt{(x_n-x_g)^2 + (y_n-y_g)^2} \,.
\end{align}
The data is filtered such, that the commissioning year of each generator matches the one of the demand time series. This is a simplification in the sense, that the register does not provide substations where the respective generator is attached to, only its geographical coordinates $x_g$ and $y_g$, such that this assignment comes with errors.

To assess the assignment errors of both electricity demand and the generation fleet, we can quantify the amount of renewable energy available at the node which cannot be consumed locally or exported due to the constraint (\ref{eq:linecapacity-constraint}), i.e. the excess which necessarily must be curtailed:
\begin{align} \label{eq:rest}
\sum_{n,\, s\in RE,\, t} \Big[ A_{n,s,t} - d_{n,t} - \sum_{\substack{\ell_{i,j} \in \mathcal{L}:\\i=n \lor j=n}} 0.7 \cdot  F_{\ell_{i,j}}^{a} \Big]^+\,.
\end{align}
The bracket $[\cdot]^+$ denotes the positive part of a value; $\max(0,\cdot)$.
Equation (\ref{eq:rest}) captures an assignment imbalance in quantities of excess TWh: If the result is positive, it indicates that the installed potential at a local substation $n$ is higher than local demand \textit{and} higher than the transmission capacity. Building such a powerplant is uneconomical, because it is known in advance that its power output cannot be used. Therefore, we assume either the assignment of $g$ to $n$ or the heuristic (\ref{eq:loadassignment}) to be inaccurate. Evaluation of (\ref{eq:rest}) can be done a priori, i.e. without solving the optimisation problem (\ref{eq:systemcosts}) with its corresponding constraints (\ref{eq:KCL})-(\ref{eq:RE-constraint}).

\subsection{Network clustering} \label{sec:clustering}

\begin{figure}[!t] \centering
	\includegraphics[width=8cm]{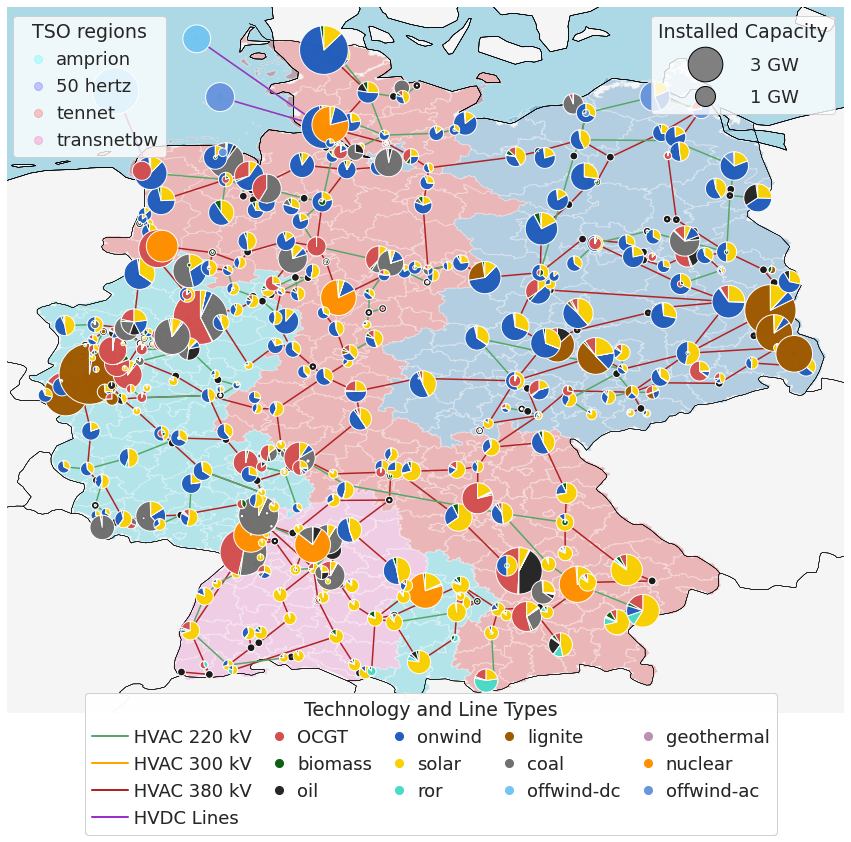}
	\caption{Original network model for Germany including all HVAC and HVDC transmission lines and all the powerplants available for the model in 2018.}
	\label{fig:germanymap}
	
	\begin{tabular} {m{0.2\textwidth} m{0.2\textwidth}}
		\begin{center} 2015 \vspace*{-.25cm} \end{center} & \begin{center} 2017 \vspace*{-.25cm} \end{center} \\ \toprule
		\includegraphics[trim=0 5.5cm 0 0, clip, width=.21\textwidth]{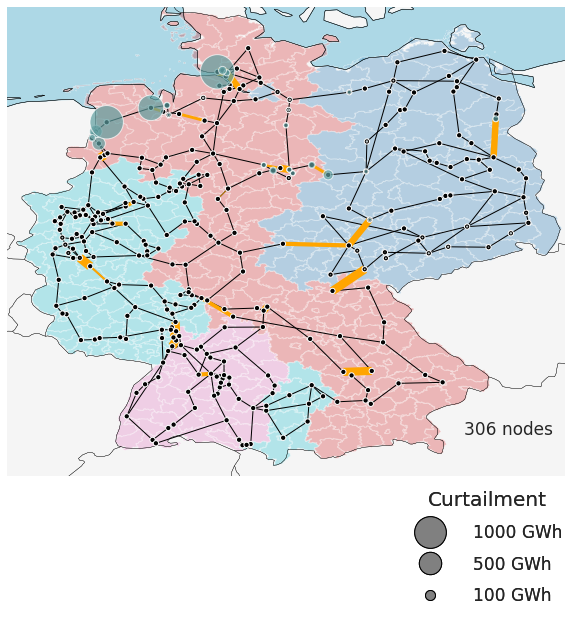}
		& \includegraphics[trim=0 5.5cm 0 0, clip, width=.21\textwidth]{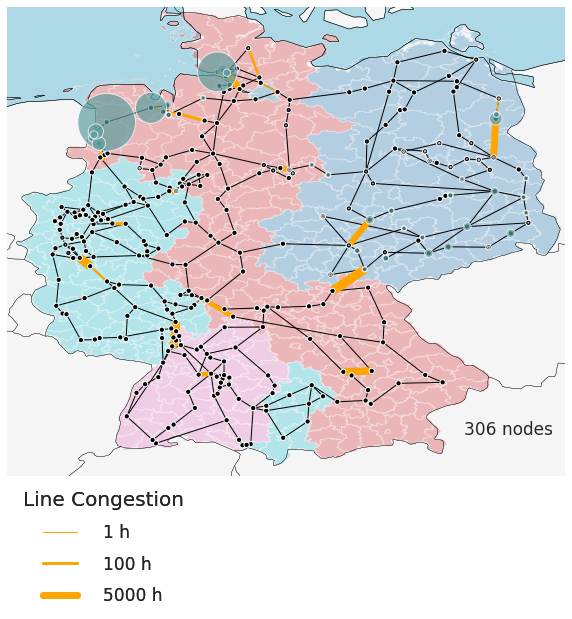} \\
		\includegraphics[width=.21\textwidth]{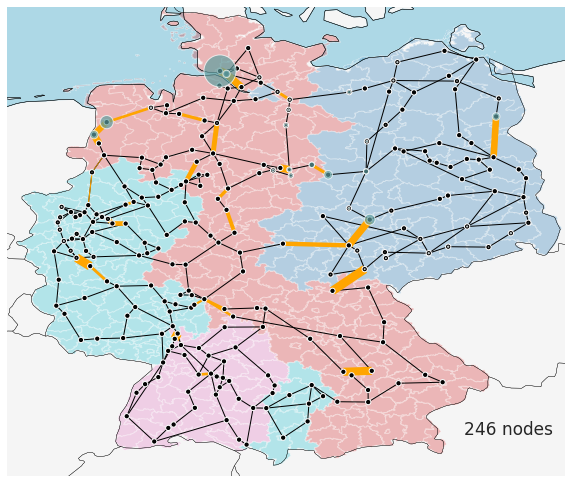}
		& \includegraphics[width=.21\textwidth]{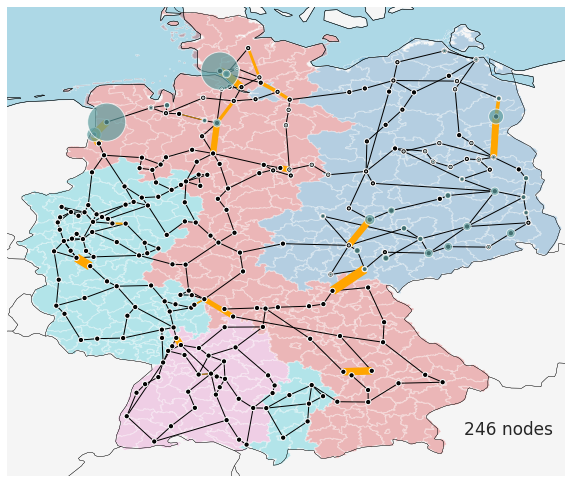} \\
		\includegraphics[width=.21\textwidth]{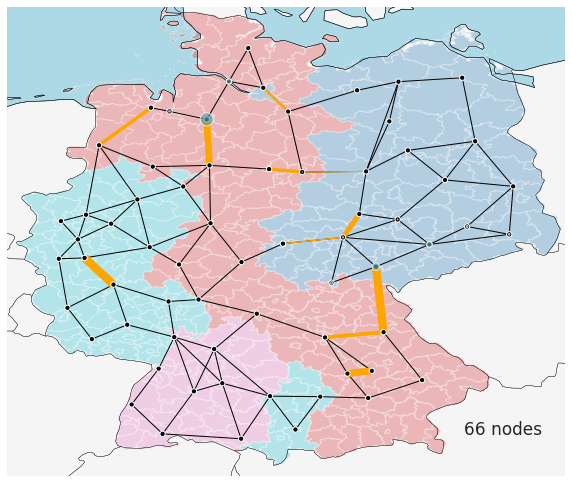}
		& \includegraphics[width=.21\textwidth]{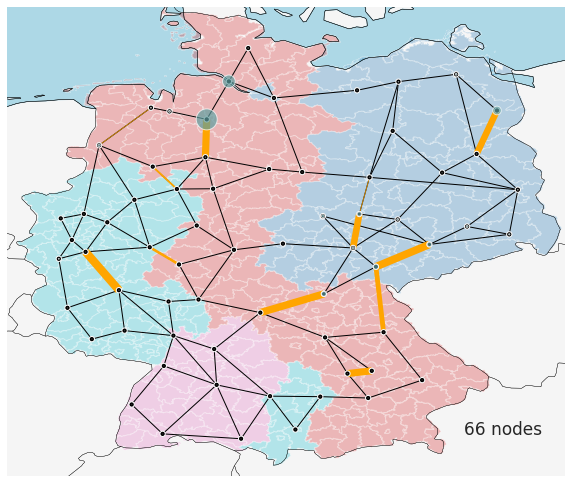} \\
		\includegraphics[trim=0 0 0 17cm, clip, width=.21\textwidth]{germany_2015_306.png}
		& \includegraphics[trim=0 0 0 17cm, clip, width=.21\textwidth]{germany_2017_306.png} \\
		\toprule
	\end{tabular}
	\caption{Clustered networks displaying the amount of curtailment for the years 2015 and 2017 for three different resolutions. The raw clustering is displayed in black, additional information for curtailment and line congestion is highlighted in blue and orange. The results for 2015 exclude i.e. the ``Thüringer Strombrücke`` that was commissioned by the end of 2015 and is hence excluded for the simulation. In 2017, the first part of the ``Thüringer Strombrücke`` is already included.} \label{fig:germanymap_clustered}
\end{figure}

\begin{figure*}[!t]
	\renewcommand{\arraystretch}{2}
	\includegraphics[width=\textwidth]{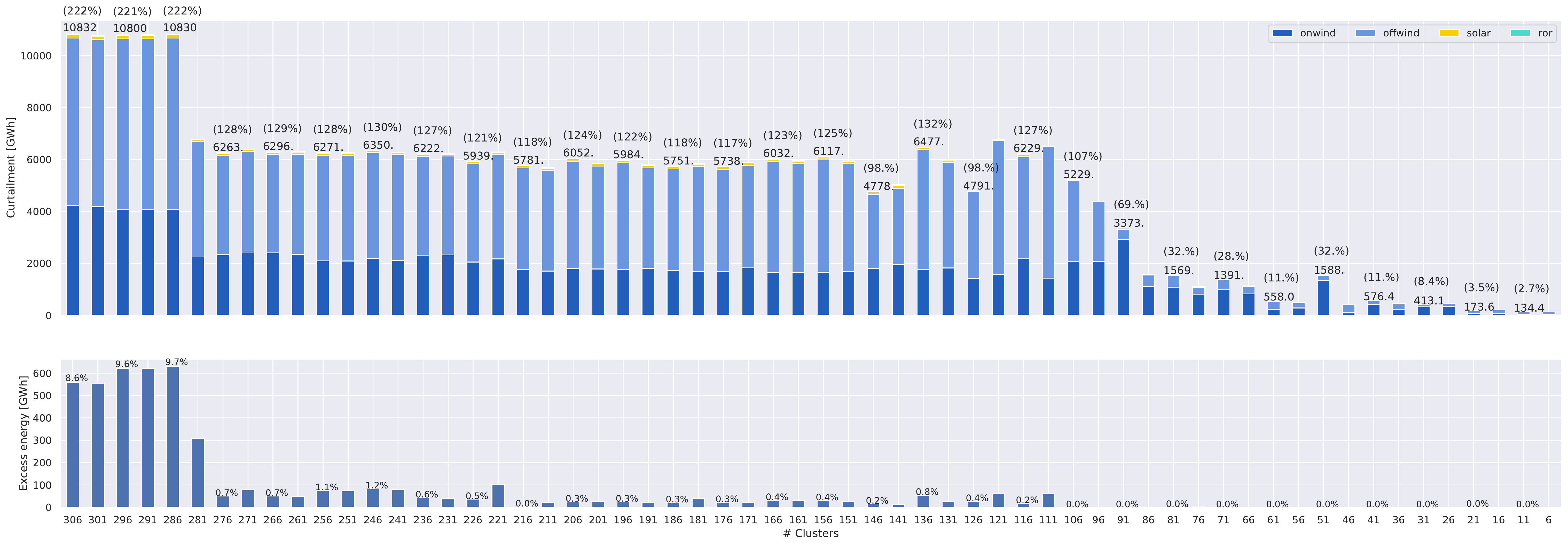}
	\caption{Model curtailment in GWh (top), excess energy according to equation (\ref{eq:rest}) in GWh (middle) and renewable availability in TWh (bottom) for the weather year 2017 respective network size interpolating between $306$ nodes (the full network) and $6$ nodes.}
	\label{fig:clustering2017}
\end{figure*}

As renewables energy carriers have gained in capacity over the past decades, more detailed models are required to detect transmission bottlenecks, thus optimisation models have grown in size, posing a great challenge to the available computational power. Many methods have been suggested to represent large models in equivalents of smaller size. To lessen the computational power on spatial scale, these methods include k-means clustering \cite{Hartigan1979}, k-means++, max-p regions \cite{Duque2012} or variants mixing different of these techniques \cite{Siala2019}. Taking into account the transmission system, clustering on electrical distances between nodes \cite{Temraz1994}-\cite{Sanchez2013} or spectral partitioning of the Laplacian matrix \cite{Hamon2015} were also developed.

We chose a version of k-means clustering based on the geographical location of the original substations in the network, weighted by the average electricity demand and conventional capacity at the substations as introduced in \cite{Hoersch2017}. This represents how the topology of the network was historically planned to connect major generators to major loads. Figure \ref{fig:germanymap_clustered} visualises the clustering for three different resolutions of the network.

After aggregating univalent nodes to their polyvalent neighbors, the k-means algorithm minimises the weighted sum of the euclidean metric
\begin{align} \label{eq:kmeans}
	\min_{x_c, y_c} \sum_{c=1}^{k}\sum_{n\in N_c} w_n \sqrt{(x_c - x_n)^2 + (y_c-y_n)^2} \,,
\end{align}
where the weight $w_n$ equals the sum of installed capacity and electricity demand at node $n$. This weight decreases the probability to aggregate nodes with high installed capacity or high electricity demand into one cluster $N_c$, such that the lines connecting nodes with high capacity and electricity demand remain isolated in the network to reflect possible transmission bottlenecks.

The node representing the cluster $N_c$ is located at the average position. All installed capacities are aggregated by technology type and demand profiles are added to the network. The time-dependent availability time series is aggregated with a weighting by technology type, such that the capacity factor of locations with high installed renewables is dominant in the average.

Lines connecting distinct clusters are represented with a single representative line with the summed capacity of all inter-cluster high-resolution lines $F_\ell$. The length of the representative line is determined using the haversine function that calculates the great-circle distance between two representative nodes and multiplied by a factor of $1.25$ to account for routing.

We progressively cluster our high resolution German model with $306$ nodes and $406$-$411$ lines (2013/2018) down to a $6$ node network and compare results.

\section{Results} \label{sec:results}

The original full-resolution network model with assigned power plants is shown in Figure \ref{fig:germanymap} and can be compared to three clustered down networks for the years 2015 and 2017 in Figure \ref{fig:germanymap_clustered}, that also shows additional spatial information of curtailment.
Annual curtailment results for different network resolutions for the year 2017 are displayed in Figure \ref{fig:clustering2017}. In \ref{sec:clusteringonresults}, we discuss a preferable network resolution for most studies and take the result as a basis for further investigation, but inter-cluster variations are discussed as well.
 
\begin{figure}[!t] \centering
	\includegraphics[width=8cm]{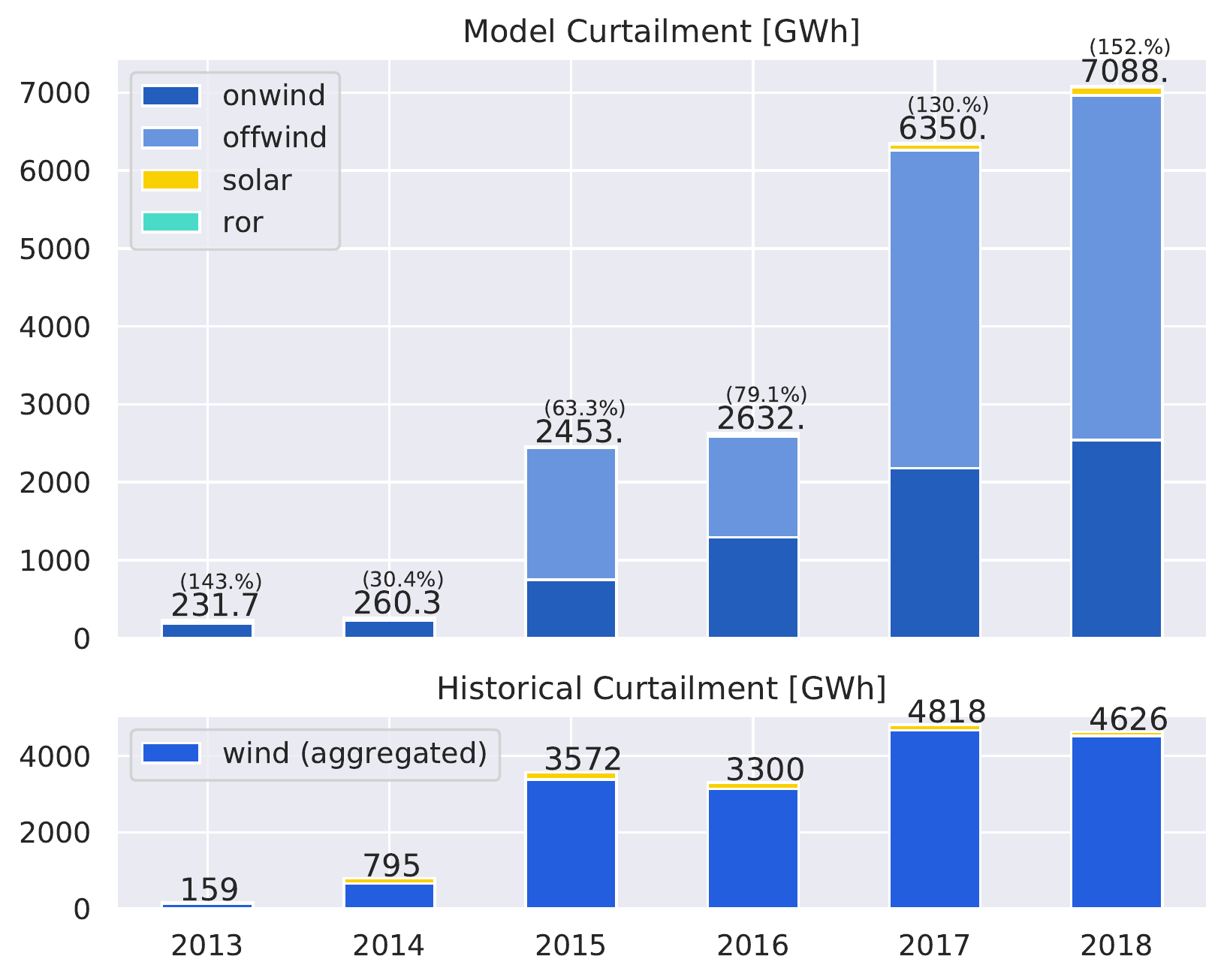}
	\caption{Model results on historical curtailment in 2013-2018 due to congestion of transmission lines in the transmission system (top). The number in percent depicts the agreement with historical data (bottom) where curtailment was caused by congestion in the transmission system \cite{EEG18}. Results are extracted at a network resolution of 246 nodes.}
	\label{fig:historicalcurtailment}
\end{figure}

Historical model results on curtailment in Germany are validated in three separate validation steps. First, on a cumulative scale where we present the total annual curtailment results in Figure \ref{fig:historicalcurtailment}. Second, on a spatial scale, where results per distinct control zone of the German transmission system operators are presented in Table \ref{tab:curtailmentpertso}. Finally, the temporal scale is considered and curtailment is validated per quarter, see Table \ref{tab:curtailmentperquartal}.

\begin{figure}
	\includegraphics[width=0.5\textwidth]{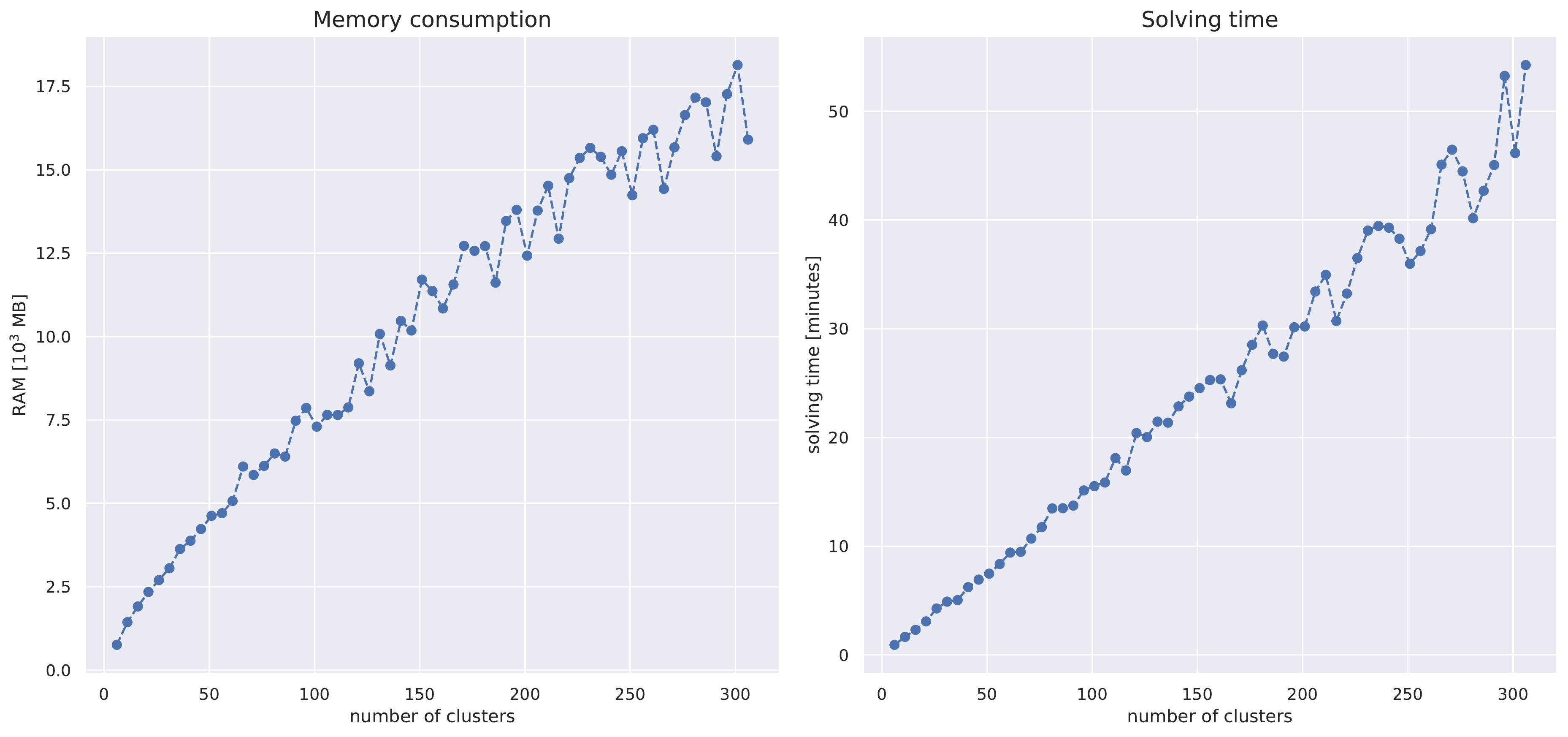}
	\caption{Memory consumption and solving time per cluster resolution, exemplary numbers for the year 2017.} \label{fig:memoryandtime}
\end{figure}

Finally, Figure \ref{fig:memoryandtime} presents memory consumption and the duration to solve the optimisation problem (\ref{eq:systemcosts})-(\ref{eq:RE-constraint}) as a function of the network size.

We have tested all results for stability by perturbing the assignment of generators $g$ according to (\ref{eq:generatorassignment}) with a probability of $5\%$ to a node that is within the radius of $31$km of the closest node $n$. $31$km account for $5\%$ of the longest east-west extension of Germany. The results are stable with deviations of below $5\%$.

\subsection{The impact of spatial resolution on modeling results} \label{sec:clusteringonresults}

In Figure \ref{fig:clustering2017} we display the impact of clustering on curtailment results in Germany in 2017. The total amount of curtailment experiences four distinct stages, where in each stage the cumulative annual value is approximately steady, deviating from the mean by only $5\%$.

Those four stages can be distinguished by applying the excess energy measure discussed in (\ref{eq:rest}): (i) First, at high model resolution, results are highly overestimated by more than $100\%$ on average. Both curtailment and excess energy results are relatively stable with minor deviations of up to $5\%$. This is because both the assignment of electricity demand and power plants according to (\ref{eq:loadassignment}) and (\ref{eq:generatorassignment}) in some cases is not precise, hence a mismatch emerges between low demand and high generation with lacking transmission capacities to transport the excess. (ii) Second, at intermediate model resolution between 250 and 150 nodes, curtailment results match the historical ones by on average $128\%$, i.e. deviating from historical numbers by $28\%$. At this stage, the effect of clustering overcomes the errors made by assigning electricity demand and generators to nodes, but at the same time, clustering preserves major transmission bottlenecks. Excess energy is still available due to the uncertainty of weather conditions, but is low at $0-1\%$ of available renewable energy. (iii) The third stage ranges from an intermediate to low resolution network ($150$ to $80$ nodes) where both curtailment and excess results have large fluctuations. This high variance is because the probability to cluster important transmission lines becomes higher as fewer clusters are available for aggregation: Similar resolutions of plus or minus 10 nodes result in different minima of the k-means objective (\ref{eq:kmeans}), and the choice of $N_c$ to preserve major bottlenecks is crucial. (iv) In the final stage, the clustering technique overcomes the transmission constraints (\ref{eq:KCL})-(\ref{eq:linecapacity-constraint}) and hence curtailment is highly underrated by $95\%\pm5\%$: Without transmission constraints, all renewable energy is consumed because of its low marginal cost.

\subsection{Cumulative results} \label{sec:cumulative}

Model results to simulate historical curtailment are shown in Figure \ref{fig:historicalcurtailment}, presenting a breakdown per carrier. It also displays the installed capacity of renewables. 

Although the marginal costs of renewables were artificially assigned for the optimisation in (\ref{eq:systemcosts}) to fix the curtailment order, the energy-mix deviates from the historical mix only by $2.5\%$ on average. We differentiate between solar, wind (onshore and offshore) and hydroelectricity.

The chosen resolution to model historical curtailment takes into account the analysis of the previous Section \ref{sec:clusteringonresults} to balance the mismatch of assigning input data to nodes versus clustering the transmission grid and overcoming important transmission bottlenecks. We choose a resolution of $246$ nodes as the excess according to (\ref{eq:rest}) is in the range of $0-10\%$ of the annual available renewable energy. Minor fluctuations are tolerated as they might not be compensated due to uncertain weather conditions. However, if excess energy accounts for more than $50\%$ of the annual available energy, it would be a highly uneconomical location of the power plant, because it is known \textit{in advance} that a large amount of the years energy can not be used.


A trend is seen that before 2017, the model tends to underestimate the total curtailment, with only 80\% of the historical curtailment captured in 2016. However as wind generation grows, the model overestimates the congestion and therefore the curtailment, reaching 50\% more than the historical numbers in 2018.

\subsection{Curtailment per TSO area} \label{sec:tso}

\begin{table}[!t] \centering
	\renewcommand{\arraystretch}{1.3}
	\caption{Curtailment per Control Zone}
	\label{tab:curtailmentpertso}
	\begin{tabular}{|c||c|c|c|c|}
		\hline
		Year&TSO zone&Historical share [\%]&Model share [\%]&Error\\
		\hline \hline
		2017 &
		50Hertz & 17.7 & 20.1 & $+1.6$ \\
		\ & TenneT & 81.6 & 80.0 & $-1.6$ \\
		\ & Transnet & 0.1 & 0 & $-0.1$ \\
		\ & amprion & 0.7 & 0 & $-0.7$ \\
		\hline
		2018 &
		50Hertz & 12 & 21.6 & $+9.6$ \\
		\ & TenneT & 87 & 78.4 & $-8.6$ \\
		\ & Transnet & 0.3 & 0 & $-0.3$ \\
		\ & amprion & 0.7 & 0 & $-0.7$ \\
		\hline
	\end{tabular}\\
	\begin{flushleft}
		Historical and model share in percent of curtailment per control zone in Germany in the years 2017 and 2018. The right column displays the model error compared to the historical data of \cite{BNetzA18}. All results are extracted from networks at a resolution of 246 nodes.
	\end{flushleft}
\end{table}

To investigate the spatial distribution of curtailment across Germany, results per control area are presented in Table \ref{tab:curtailmentpertso} for the years 2017 and 2018 in percent of annual curtailment. For consistency, the model resolution is chosen such as in Section \ref{sec:cumulative}. Results indicate, that curtailment numbers in our model deviate by up to $9\%$ from historical values, and by $3.6\%$ on average.

A validation of how these result change with the number of nodes representing the model show the same four stages as discussed in Section \ref{sec:clusteringonresults}: In stage (i), where curtailment results are highly overestimated and electricity demand and generators were assigned to incorrect nodes, curtailment in 2017 is split $92:8$ between TenneT and 50Hertz,  $87:12$ in 2018. These numbers deviate in both years by $\pm 1\%$ as the network resolution changes. The balancing of stage (ii) results in a stable $80:20$ split between TenneT and 50Hertz in 2017, and $78:22$ in 2018 with deviations of up to $2\%$ in both years as the network resolution changes. Stage (iii) remains relatively random, which is true for stage (iv) as well, but in the latter, total annual curtailment is so low, such that the proportionality has no meaning.

\subsection{Curtailment per Quarter} \label{sec:quarterly}

\begin{table}
	\renewcommand{\arraystretch}{1.3}
	\caption{Curtailment per Quarter}
	\label{tab:curtailmentperquartal}
	\begin{tabular}{|c||c|c|c|c|}
		\hline
		Year & Quarter & Historical share [\%]& Model share [\%]& Error\\
		\hline \hline
		2015 & I & 24.0 & 33.4 & +9.4 \\
		\ & II & 15.6 & 10.6 & -5.0 \\
		\ & III & 17.3 & 14.2 & -3.1 \\
		\ & IV & 43.1 & 41.8 & -1.3 \\ \hline
		
		2016 & I & 40.7 & 48.2 & +7.5 \\
		\ & II & 14.3 & 6.3 & -8.0 \\
		\ & III & 14.7 & 8.3 & -6.4 \\
		\ & IV & 30.3 & 37.2 & +7.2 \\ \hline
		
		2017 & I & 25.6 & 23.3 & -2.3 \\
		\ & II & 24.7 & 19.7 & -4.0 \\
		\ & III & 7.9 & 8.7 & +0.8\\
		\ & IV & 41.8 & 48.3 & +6.5 \\ \hline
		
		2018 & I & 36.5 & 39.6 & +3.1 \\
		\ & II & 17.5 & 13.0 & -4.5 \\
		\ & III & 13.4 & 11.1 & -2.3 \\
		\ & IV & 32.6 & 36.3 & +3.7 \\
		\hline
	\end{tabular}\\
	\begin{flushleft}
		Historical and model shares per quarter in percent of curtailment for the years 2015-2018. The right column displays the model error compared to the historical data of \cite{BNetzA19}. All results are extracted from networks at a resolution of 246 nodes.
	\end{flushleft}
\end{table}

Finally, we consider curtailment per quarter in the years 2015-2018. Results are displayed in Table \ref{tab:curtailmentperquartal} in percent of annual curtailment. The model resolution is chosen the same as in Section \ref{sec:cumulative} for reasons of consistency. Results indicate that the distribution of curtailment at an hourly resolution model reflect the historical distribution with an error of in average $4.7\%$.

Again, these results are validated on how they change with the cluster size. Here, a positive trend can be observed as the clustering happens mainly in space and not in time, such that results are stable for stages (i)-(iii). The average share is reflects the number in Table \ref{tab:curtailmentperquartal}, while it starts fluctuating towards stage (iv), where we know that the overall curtailment tends to $0$, such that the proportionality has no meaning.

\subsection{Memory consumption and Solving times}


Solving the full optimisation problem with a full resolution network of $306$ nodes and more than $400$ HVAC or HVDC lines requires approximately $20$ GB Random-Access-Memory (RAM) and runs for almost an hour, while a clustered network of only $156$ nodes is twice as fast and needs about $30\%$ less RAM, while providing more accurate results.

\section{Critical appraisal}
This case-study covers Germany only, neglecting the fact that power can also be exchanged with bordering countries, such as France, Denmark, Poland, the Czech Republic, Austria or Switzerland, reducing the overall curtailment of renewables in Germany. Previous studies have pointed out, that international cooperation benefits renewable electricity markets \cite{Schlachtberger2017}.

Further, to avoid the difficulty of keeping track of different voltage levels as the network is reduced, all lines are mapped to their electrical equivalents at 380~kV, the most prevalent voltage in the German transmission system. The electrical parameters and capacities of the lines use standard assumptions for $380$ kV circuits whereas in reality they vary from line to line. In addition, we use constant summer thermal ratings for an outside temperature of 20 Celsius throughout the year and do not adapt them for lower temperatures or wind conditions. The use of winter ratings as well as dynamic line rating \cite{Wallnerstroem2015} on some congested lines today may account for the lower historical curtailment compared to our model. 

Finally, the capacity factors for wind and solar from weather data overestimate historical production, so we linearly reduced them by a factor of 0.9 for wind and 0.8 for solar for each point in time and space.

\section{Discussion and Conclusions}
We have shown that historic curtailment in Germany can be reproduced in the open model PyPSA-Eur using the latest database for the locations of existing renewable
generators. Results agree well in time (curtailment per quarter) and space (curtailment per TSO region), provided a balancing resolution is used that is low enough to overcome assignment-errors and high enough to account for important transmission routes. We suggest to cluster the 306 node network to well below 280 nodes, but not below 150 nodes. In this range, curtailment in the model provides the best match with historical data. A resolution below 100 nodes for Germany using a weighted k-means clustering scheme is not advisable.

\section*{Acknowledgments}
We thank Fabian Neumann, Elisabeth Zeyen and Frederick Unnewehr for helpful discussions, suggestions and comments.

\end{document}